\documentclass[12pt]{article}

\usepackage{amsbsy}
\usepackage{amsfonts}
\usepackage{amssymb}

\usepackage{bm}
\usepackage{enumerate}
\usepackage{graphicx}   
\usepackage{epsfig}
\usepackage{color}
\usepackage{nicefrac}
\usepackage{mathrsfs}
\usepackage{amsmath}
\usepackage{amsthm}
\usepackage{subcaption}

\DeclareFontFamily{OT1}{pzc}{}
\DeclareFontShape{OT1}{pzc}{m}{it}{<-> s * [1.200] pzcmi7t}{}
\DeclareMathAlphabet{\mathpzc}{OT1}{pzc}{m}{it}
\input epsf.sty

\newlength{\TZ}
\newcommand{\BEQ}{\begin{equation}}     
\newcommand{\BEA}{\begin{eqnarray}}
\newcommand{\BD}{\begin{displaymath}}
\newcommand{\EEQ}{\end{equation}}       
\newcommand{\EEA}{\end{eqnarray}}
\newcommand{\ED}{\end{displaymath}}
\newcommand{\vep}{\varepsilon}          
\newcommand{\vph}{\varphi}              
\newcommand{\D}{{\rm d}}                
\newcommand{\II}{{\rm i}}               
\newcommand{\demi}{\frac{1}{2}}         
\newcommand{\wht}[1]{\widehat{#1}}      

\renewcommand{\vec}[1]{\boldsymbol{#1}} 

\newcommand{\appsection}[2]{\setcounter{equation}{0}\setcounter{subsection}{0}
\section*{Appendix #1. #2}
\renewcommand{\theequation}{#1.\arabic{equation}}
              \renewcommand{\thesection}{#1} }

\catcode`\@=11
\def\numberbysection{\@addtoreset{equation}{section}
        \def\theequation{\thesection.\arabic{equation}}}
\numberbysection

\definecolor{gruen}{rgb}{0,0.625,0}     
\definecolor{rot}{rgb}{0.75,0,0}        
\definecolor{blau}{rgb}{0,0,0.75}       
\newcommand{\BLAU}[1]{\textcolor{black}{{\rm #1}}}	

\begin{document}

\begin{titlepage}

\vskip 1.5 cm
\begin{center}
{\LARGE \bf Boundedness of meta-conformal two-point functions in one and two spatial dimensions}
\end{center}

\vskip 2.0 cm
\centerline{{\bf Malte Henkel}$^{a,b,c}$, {\bf Michal Dariusz Kuczynski}$^a$\footnote{present address: 
Max-Planck Institut f\"ur Plasmaphysik, Wendelsteinstra{\ss}e 1, D -- 17491 Greifswald, Germany} and {\bf Stoimen Stoimenov}$^d$}
\vskip 0.5 cm
\centerline{$^a$ Laboratoire de Physique et Chimie Th\'eoriques (CNRS UMR 7019), Universit\'e de Lorraine Nancy,}
\centerline{B.P. 70239, F -- 54506 Vand{\oe}uvre l\`es Nancy Cedex, France\footnote{permanent address}}
\vspace{0.5cm}
\centerline{$^b$ Centro de F\'{i}sica Te\'{o}rica e Computacional, Universidade de Lisboa, P -- 1749-016 Lisboa, Portugal}
\vspace{0.5cm}
\centerline{$^c$ Max-Planck Institut f\"ur Physik Komplexer Systeme, N\"othnitzer Stra{\ss}e 38, D -- 01187 Dresden, Germany}
\vspace{0.5cm}
\centerline{$^d$ Institute of Nuclear Research and Nuclear Energy, Bulgarian Academy of Sciences,}
\centerline{72 Tsarigradsko chaussee, Blvd., BG -- 1784 Sofia, Bulgaria}

\begin{abstract}
Meta-conformal invariance is a novel class of dynamical symmetries, with dynamical exponent $z=1$, and distinct from the
standard ortho-conformal invariance. The meta-conformal Ward identities can be directly read off from the Lie algebra
generators, but this procedure implicitly assumes that the co-variant correlators should depend holomorphically on time-
and space coordinates. Furthermore, this assumption implies un-physical singularities in the co-variant
correlators. A careful reformulation of the global meta-conformal Ward identities in a dualised space, 
combined with a regularity postulate, leads to 
bounded and regular  expressions for the co-variant two-point functions, both in $d=1$ and $d=2$ spatial dimensions.
\end{abstract}
\end{titlepage}

\setcounter{footnote}{0}

\section{Introduction}

Dynamical symmetries have been an extremely useful tool in the analysis of strongly interacting many-body systems. 
Probably the most striking example are the (ortho-)conformal transformations in $(1+1)$-dimensional 
time-space \cite{Belavin84}, for textbooks see \cite{Francesco97,Henkel99} \BLAU{or \cite{Schottenloher08,Blumenhagen09,HenkelKarevski12}}. 
Expressed in orthogonal light-cone coordinates $z=t+\II r$, $\bar{z}=t-\II r$, these
transformations are isomorphic to the direct product of holomorphic transformations $f(z)$ 
and anti-holomorphic transformation $\bar{f}(\bar{z})$, 
see table~\ref{tab1}, whose Lie algebra is the direct sum of two Virasoro algebras. 
For ortho-conformal invariance, a convenient Lie algebra basis is spanned
by the generators (with $j\in\mathbb{Z}$)
\BLAU{\BEQ \label{1.1}
\ell_j = -z^{j+1}\partial_z - (j+1)\Delta z^j \;\; , \;\;
\bar{\ell}_j = -\bar{z}^{j+1}\partial_{\bar{z}} - (j+1)\overline{\Delta} \bar{z}^j 
\EEQ} 
\hspace{-0.18truecm}where the ortho-conformal weight $\Delta$ 
describes the transformation of the associated scaling operator $\phi$ on which $\ell_j$ acts (and similarly $\bar{\ell}_j$ 
for the variable $\bar{z}$ with the weight $\overline{\Delta}$). \BLAU{Since the ortho-conformal Lie algebra includes
the generator $\ell_0-\bar{\ell}_0$ of rotations, ortho-conformally invarianct systems are 
{\it a fortiriori} rotation-invariant. The most simple example of this is the conformally invariant $(1+1)D$ Laplace
equation $\partial_z \partial_{\bar{z}}\phi(z,\bar{z})=0$.}

The one-particle generators $\ell_j$ can be lifted to $N$-particle generators $\ell_j^{[N]} = \sum_{p=1}^{N} \ell_j^{(p)}$, 
where $\ell_j^{(p)}$ gives the action of the generator $\ell_j$ on the $p^{\rm th}$ scaling operator 
$\phi_p=\phi_p(z_p,\bar{z}_p)$. 
One of the most immediate applications of such symmetries is the derivation of co-variant $N$-point functions 
$C^{[N]} = C^{[N]}(z_1,\bar{z}_1,\ldots,z_N,\bar{z}_N) 
= \left\langle \phi_1(z_1,\bar{z}_1) \ldots \phi_N(z_N,\bar{z}_N)\right\rangle$ 
built from physical scaling operators $\phi_i(z_i,\bar{z}_i)$, $i=1,\ldots,N$. 
The co-variance of the $N$-point function is then expressed through a closed set of linear first-order 
differential equations, \BLAU{the ortho-conformal Ward identities \cite{Belavin84}.} 
\BLAU{For $N=2$ or $N=3$, it is enough to consider merely the maximal finite-dimensional
Lie sub-algebra, when these equations are of the form $\ell_{j}^{[N]} C^{[N]}=0$ (with $j=-1,0,1$). Their} 
solutions are holomorphic in the variables $z_p$ (or anti-holomorphic in the variables 
$\bar{z}_p$) \cite{Hille76}. For example, the ortho-conformal two-point function
has the well-known form ($C_0$ is a normalisation constant) 
\BEQ \label{1}
C_{\rm ortho}^{[2]}(z_1,\bar{z}_1,z_2,\bar{z}_2) 
= C_0 \, \delta_{\Delta_1,\Delta_2} \delta_{\overline{\Delta}_1,\overline{\Delta}_2} \:
\bigl( z_1 - z_2\bigr)^{-2\Delta_1} \bigl( \bar{z}_1 - \bar{z}_2 \bigr)^{-2\overline{\Delta}_1}
\EEQ

\begin{figure}[b]
\begin{center}
\includegraphics[width=.475\hsize]{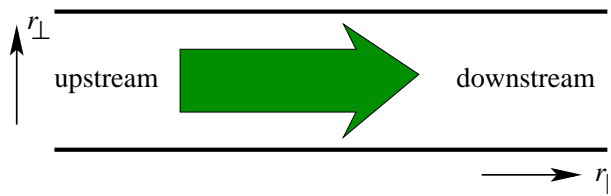} ~~~
\end{center}
\caption[fig0]{Schematic illustration of ballistic transport in a channel, 
with the spatial coordinates $r_{\|}$, $r_{\perp}$. The directional bias is indicated. \label{fig0}}
\end{figure}

{\small \begin{table}[tb]
\hspace{-0.5truecm}\begin{tabular}{|l|lll|l|c|l|} \hline
group                    & \multicolumn{3}{l|}{coordinate changes} & phys. coordinates & $\mathpzc{z}$ & co-variance \\ \hline
ortho-conformal $(1+1)D$ & $z'=f(z)$       & $\bar{z}'=\bar{z}$           &       & $z = t+\II \mu r$   & $1$ & correlator  \\
                         & $z'=z$          & $\bar{z}'=\bar{f}(\bar{z})$  &       & $\bar{z}=t-\II \mu r$  &  & \\ \hline
meta-conformal $1D$      & $u=f(u)$        & $\bar{u}'=\bar{u}$           &       & $u=t$               & $1$ & correlator \\
                         & $u'=u$          & $\bar{u}'=\bar{f}(\bar{u})$  &       & $\bar{u}=t+\mu r$   &     & \\ \hline
meta-conformal $2D$      & $\tau'=b(\tau)$ & $w'=w$    & $\bar{w}'=\bar{w}$       & $\tau=t$            &     & \\
                         & $\tau'=\tau$    & $w'=f(w)$ & $\bar{w}'=\bar{w}$       & $w=t+\mu (r_{\|}+\II r_{\perp})$ 
                                                                                                        & $1$ & correlator \\
                         & $\tau'=\tau$    & $w'=w$ & $\bar{w}'=\bar{f}(\bar{w})$ & $\bar{w} = t +\mu(r_{\|}-\II r_{\perp})$ 
                                                                                                        &     & \\ \hline
conformal galilean       & $t'=b(t)$       & \multicolumn{2}{l|}{$\vec{r}'=\left(\D b(t)/\D t\right) \vec{r}$} & $t$ &  &  \\
                         & $t'=t$          & $\vec{r}'=\vec{r}+\vec{a}(t)$  &     & $\vec{r}$           & $1$ & correlator \\ 
                         & $t'=t$          & $\vec{r}'=\mathscr{R}(t)\vec{r}$ &   &                     &     & \\ \hline
Schr\"odinger-Virasoro   & $t'=b(t)$       & \multicolumn{2}{l|}{$\vec{r}'=\left(\D b(t)/\D t\right)^{1/2}\vec{r}$} & $t$ & &\\
                         & $t'=t$          & $\vec{r}'=\vec{r}+\vec{a}(t)$  &     & $\vec{r}$  & $2$    & response \\ 
                         & $t'=t$          & $\vec{r}'=\mathscr{R}(t)\vec{r}$ &   &            &        & \\ \hline
\end{tabular}
\caption[tab1]{
Several infinite-dimensional groups of time-space transformations, defined by the corresponding coordinate changes. 
Unspecified (vector) functions are assumed (complex) differentiable and 
$\mathscr{R}(t)\in\mbox{\sl SO}(d)$ is a smoothly time-dependent rotation matrix. 
The physical time- and space-coordinates, the associated dynamical exponent $\mathpzc{z}$ of this standard representation 
and the physical nature of the co-variant $n$-point functions is also indicated. 
\label{tab1}}
\end{table}}

In this work, we are interested in new types of (meta-)conformal invariance \cite{Henkel02,Henkel19a,Stoimenov19} 
where the implicit hypothesis of holomorphy in the 
physical coordinates $z_p,\bar{z}_p$ (or equivalently $t_p,r_p$, with $p=1,\ldots, N$) 
is not necessarily satisfied. {\em Meta-conformal invariance} arises as a dynamical symmetry of the simple equation 
${\cal S}\vph(t,\vec{r})= \left( - \mu\partial_t + \partial_{r_{\|}}\right)\vph(t,\vec{r})=0$ 
of ballistic transport, which distinguishes a
single preferred direction \cite{Stoimenov15},  with coordinate $r_{\|}$, 
from the transverse direction(s), with coordinate $\vec{r}_{\perp}$. 
This is sketched in figure~\ref{fig0}.  A preferred spatial direction is incompatible \BLAU{with rotation-invariance, 
hence {\it a fortiriori} incompatible} with ortho-conformal invariance. 
A simple example from statistical physics of a system with a dynamical meta-conformal
invariance is provided by the kinetic Glauber-Ising model 
with a bias and long-ranged initial conditions \cite{Henkel19a,Stoimenov21}. 
Ballistic transport occurs in innumerable closed quantum systems, see e.g. 
\cite{Biella19,Bluhm19,Calabrese07,Calabrese16,Castro16,Caux17,Delfino17,Doyon17,Dutta15,Hansson17,Kang18,Piroli17} 
as well as in classical non-equilibrium dynamics, see e.g. 
\cite{Trizac02,Coppex03,Campa14,Elskens14,Godreche15b,Stoimenov15,Weiss15,Alonso18}. 
\BLAU{In table~\ref{tab1} several examples of time-space coordinate transformations are listed, including ortho- and
meta-conformal transformations. It can be seen
that $(1+1)D$ ortho-conformal transformations and $1D$ meta-conformal transformations have the same
underlying abstract algebraic structure, namely isomorphic Lie algebras, see \cite{Schottenloher08}. However, they refer to different physical choices
which represent the time-space variables.}
As \BLAU{is further} shown in table~\ref{tab1}, for $d=1$ and $d=2$ space dimensions, 
the Lie group of meta-conformal transformations is infinite-dimensional. 
The Lie algebra generators of meta-conformal invariance are read off from table~\ref{tab1} as follows, \BLAU{in terms of the 
physical time-space coordinates $t$ and $r$ (respectively $r_{\|}, r_{\perp}$).} 
In the $1D$ case, in terms of time- and space-coordinates \cite{Henkel02} (with $n\in\mathbb{Z}$)
\BEA
\ell_n &=& -t^{n+1}\left( \partial_t - \frac{1}{\mu}\partial_r\right) - (n+1) \left( \delta - \frac{\gamma}{\mu} \right) t^n 
\nonumber \\
\bar{\ell}_n &=& -\frac{1}{\mu}\bigl(t+\mu r\bigr) \partial_r - (n+1) \frac{\gamma}{\mu} \bigl(t+\mu r\bigr)^n
\label{3}
\EEA
\BLAU{(notice the differences with respect to (\ref{1.1}))} and in the $2D$ case \cite{Henkel19a}
\BEA
A_n &=& -t^{n+1}\left( \partial_t - \frac{1}{\mu}\partial_{\|}\right) 
        -(n+1)\left(\delta - \frac{2\gamma_{\|}}{\mu}\right) t^n  \label{4} \\
B_n^{\pm} &=& -\frac{1}{2\mu}\bigl( t+\mu(r_{\|}\pm\II r_{\perp})\bigr)^{n+1} \bigl(\partial_{\|}\mp\II\partial_{\perp}\bigr) 
-(n+1)\frac{\gamma_{\|}\mp\II\gamma_{\perp}}{\mu} \bigl( t+\mu(r_{\|}\pm\II r_{\perp})\bigr)^n
\nonumber
\EEA
with the short-hands $\partial_{\|}=\frac{\partial}{\partial r_{\|}}$ and 
$\partial_{\perp}=\frac{\partial}{\partial r_{\perp}}$. 
The constants $\delta$ and $\gamma$ (respectively $\gamma_{\|,\perp}$) are the scaling dimension and the rapidity of the
scaling operators on which these generators act and $\mu^{-1}$ is a constant with the dimension of a velocity. 
Each of the infinite families of generators in (\ref{3},\ref{4}) produces a Virasoro algebra (with zero central charge).
Therefore, the $1D$ meta-conformal algebra is isomorphic to a direct sum of two Virasoro algebras. In the $2D$ case,
there is an isomorphism with the direct sum of three Virasoro algebras. Their maximal finite-dimensional Lie sub-algebras
(isomorphic to a direct sum of two or three $\mathfrak{sl}(2,\mathbb{R})$ algebras) fix the form of two-point
correlators $C(t,\vec{r}) = \left\langle \vph_1(t,\vec{r}) \vph_2(0,\vec{0}) \right\rangle$ 
built from quasi-primary scaling operators. 
Since the generators (\ref{3},\ref{4}) already contain the terms which describe how the scaling operators 
$\vph=\vph(t,\vec{r})$ transform under their action, the global meta-conformal Ward identities can simply be written down.
The requirement of meta-conformal co-variance leads to \cite{Henkel19a}
\BEA
\lefteqn{C_{\rm meta}^{[2]}(t,\vec{r}) = \left\langle \vph(t,\vec{r}) \vph(0,\vec{0}) \right\rangle} \label{1.4} \\
&=& 
\left\{ 
\begin{array}{ll} 
\delta_{\delta_1,\delta_2} \delta_{\gamma_1,\gamma_2} C_0\, 
t^{-2\delta_1} \left( 1 +  \mu \frac{r}{t} \right)^{-2\gamma_1/\mu} & \mbox{\rm ~;~ if $d=1$} \\
\delta_{\delta_1,\delta_2} \delta_{\gamma_{\|,1},\gamma_{\|,2}}\delta_{\gamma_{\perp,1},\gamma_{\perp,2}} C_0\, 
t^{-2\delta_1} \left( 1 + {\mu} \frac{r_{\|}+\II r_{\perp}}{t} \right)^{-2\gamma_1/\mu} 
\left( 1 + {\mu} \frac{r_{\|}-\II r_{\perp}}{t} \right)^{-2\bar{\gamma}_1/\mu} & \mbox{\rm ~;~ if $d=2$}
\end{array} \right. 
\nonumber
\EEA
and where $\vec{r}=r\in\mathbb{R}$ for $d=1$ and $\vec{r}=(r_{\|},r_{\perp})\in\mathbb{R}^2$ for $d=2$. 

\begin{figure}[ht]                 
\begin{subfigure}{0.5\textwidth}   
   \centering
   \includegraphics[width=.85\hsize]{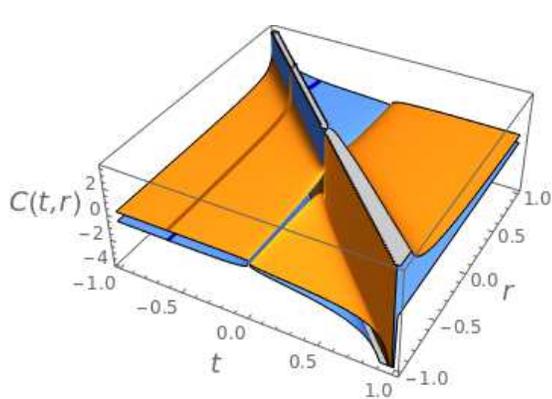}
   \caption[fig00a]{Spurious singularities arise in eq.~(\ref{1.4}).}
   \label{fig:sub-first}
\end{subfigure}
\begin{subfigure}{0.5\textwidth}
   \centering
   \includegraphics[width=.85\hsize]{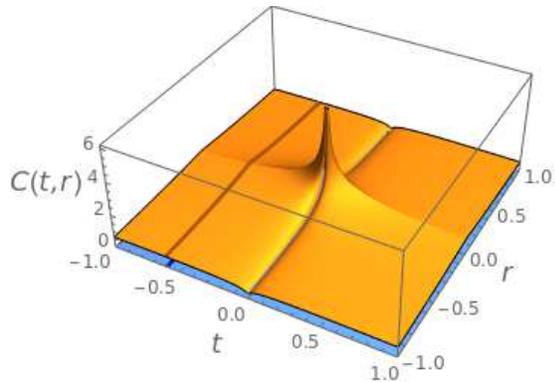}
   \caption{Regularised bounded real-valued form eq.~(\ref{metareg1d}).}
   \label{fig:sub-second}
\end{subfigure}
\begin{subfigure}{.475\textwidth}
   \centering
   \includegraphics[width=.85\hsize]{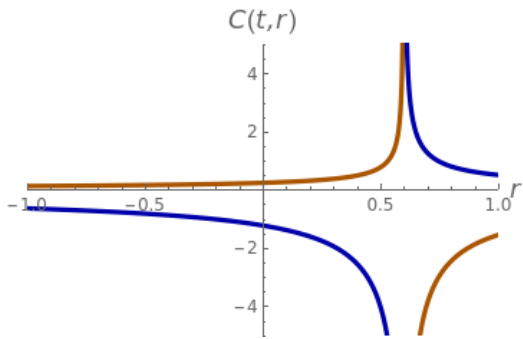}
   \caption{For $t=-0.6$, a complex-valued singularity occurs at $r=-t=0.6$ in eq.~(\ref{1.4}).}
   \label{fig:sub-third}
\end{subfigure}
\begin{subfigure}{.475\textwidth}
   \centering
   \includegraphics[width=.85\hsize]{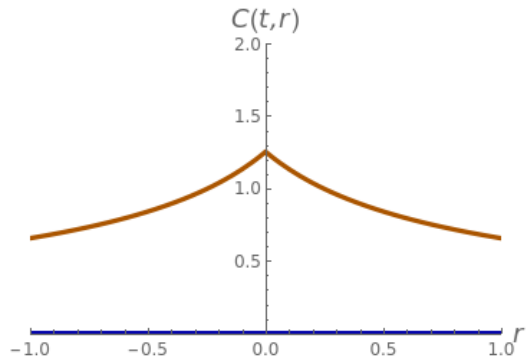}
   \caption{For $t=-0.6$, bounded real-valued behaviour throughout in eq.~(\ref{metareg1d}).}
   \label{fig:sub-forth}
\end{subfigure}
\caption[fig00]{Real part (orange) and imaginary part (blue) of the  $1D$ meta-conformally co-variant 
two-point function $C(t,r)$, with $\delta_1=0.22$, $\gamma_1=0.33$ and $\mu_1=\mu_2=1$. 
\BLAU{The dark straight lines in figure~\ref{fig00}(a,b) indicate the section $t=-0.6$ 
along which the correlator is shown again in figure~\ref{fig00}(c,d).}
\label{fig00}}
\end{figure}

Formally, the procedure to derive (\ref{1.4}) is completely analogous to the used above for the derivation of (\ref{1}) from 
ortho-conformal co-variance. 
The explicit forms (\ref{1.4}) make it apparent that $C_{\rm meta}(t,\vec{r})$ is not necessarily 
bounded for all $t$ or $\vec{r}$. In figure~\ref{fig00}, 
we illustrate this for the $1D$ case --  a spurious singularity appears whenever ${\mu}r=-t$.

In the limit $\mu\to 0$, the meta-conformal algebras contract into the {\em galilean conformal algebras} \cite{Havas78}. 
Carrying out this limit
on the correlator (\ref{1.4}), one obtains, as has been stated countless times in the literature, see e.g.
\cite{Bagchi09,Bagchi10,Bagchi13b,Bagchi17,Martelli10}
\BEQ \label{1.5} 
C_{\mbox{\sc\footnotesize cga}}^{[2]}(t,\vec{r}) = \delta_{\delta_1,\delta_2} \delta_{\vec{\gamma}_1,\vec{\gamma}_2} \left\{ 
\begin{array}{ll}  
t^{-2\delta_1} \exp\left(-2 \frac{\gamma_1 r}{t}\right)                  & \mbox{\rm ~;~~ if $d=1$} \\[0.12truecm] 
t^{-2\delta_1} \exp\left(-4 \frac{\vec{\gamma}_1\cdot\vec{r}}{t}\right)  & \mbox{\rm ~;~~ if $d=2$}
\end{array} \right. 
\EEQ
with the definition $\vec{\gamma}=(\gamma_{\|},\gamma_{\perp})$. 
While this correlator decays in one spatial direction (`downstream', where $\gamma_1 r>0$  
or $\vec{\gamma}_1\cdot \vec{r}>0$, respectively 
and assuming $t>0$), it diverges in the opposite (`upstream') direction, as defined in figure~\ref{fig0}. 
Again, such a behaviour does not appear to be physical. In view of the large interest devoted to conformal galilean
field-theory, see \cite{Bondi62,Barnich07,Barnich13,Aizawa16,Hosseiny10b,Martelli10,Duval09,Duval14a,Duval14b,Kriv16,Bagchi17} 
and refs. therein, it appears
important to be able to formulate well-defined correlators which remain bounded everywhere in time-space. 

We mention in passing that the $1D$ form of (\ref{1.5}) can also be formally obtained from $2D$ ortho-conformal invariance: it
is enough to consider complex conformal weights $\Delta=\demi\left(\delta-\II\gamma/\mu\right)$ and
$\overline{\Delta}=\demi\left(\delta+\II\gamma/\mu\right)$. Then (\ref{1}) can be rewritten as
\BEQ \label{1.6}
C_{\rm ortho}^{[2]}(t,r) 
= t^{-2\delta}\left[ 1 +\left(\frac{\mu r}{t}\right)^2\right]^{-\delta} 
\exp\left[-\frac{2\gamma}{\mu}\arctan \frac{\mu r}{t} \right]
\stackrel{\mu\to 0}{\longrightarrow} t^{-2\delta} e^{-2\gamma r/t}
\EEQ

In what follows, we shall describe a mathematically well-defined procedure 
how to find meta-conformal correlators bounded everywhere. 
Since the implicit assumption of holomorphicity
in the coordinates gave the unbounded results (\ref{1.4},\ref{1.5}), we shall explore how to derive non-holomorphic 
correlators. 
\BLAU{The first case where this question could be treated was the one of the {\em conformal galilean algebra}, 
where the regularised form reads in $d\geq 1$ dimensions \cite{Henkel15,Henkel15a}
\BEQ \label{1.8}
C_{\rm CGA, reg}^{[2]}(t,\vec{r}) = 
\delta_{\delta_1,\delta_2} \delta_{\vec{\gamma}_1,\vec{\gamma}_2}\, 
|t|^{-2\delta_1} \exp\left( -2 \alpha \left|\frac{\vec{\gamma}_1\cdot\vec{r}}{t}\right|\:\right) F_0(\vec{\gamma}_1^2)
\EEQ
instead of (\ref{1.5}), where the function $F_0$ remains undetermined \BLAU{and $\alpha=\alpha(d)$ is a possibly dimension-dependent constant.} 
The derivation is briefly reproduced in appendix~A for $d=1$, for the convenience of the reader. In particular, 
this example shows (i) the importance of the dualisation technique to be explained in section~2 and (ii) the relevance of semi-infinite representation theorems
derived from the theory of Hardy spaces \cite{Akhiezer88,Stein71}, see appendix~B.} 
Our treatment of the meta-conformal case follows \cite{Henkel16}, to be generalised to the case $d=2$ where necessary. This in turn
is inspired by methods developed for the Schr\"odinger algebra \cite{Henkel03a,Henkel14a}. 

Throughout, we shall admit rotation-invariance in the transverse directions, if
applicable. Therefore, in more than three spatial dimensions, the consideration of the two-point function can be reduced
to the case of a single transverse direction, $r_{\perp}$. Therefore, it should be enough to discuss explicitly either (i) the
case of one spatial dimension, referred from now one as the {\em $1D$ case} (without a transverse direction), or
else (ii) the case of two spatial dimensions, called the {\em $2D$ case} (with a single transverse direction). 

This work is organised as follows. Since the  straightforwardish implementation of 
the global meta-conformal Ward identities leads 
to un-physical singularities in the time-space behaviour of such correlators, which arise since the meta-conformally 
co-variant correlators are no longer holomorphic functions of their arguments, we shall present a more careful approach  
in sections~2 and~3. These cover, respectively, $d=1$ and $d=2$ spatial dimensions. 
Technically, this proceeds via a dualisation and a regularity requirement in dual space. 
Our main result is the explicit form of a meta-conformally co-variant two-point
function which remains bounded everywhere, as stated in eqs.~(\ref{5.1},\ref{5.2}) in section~4.
\BLAU{Appendix~A outlines the construction for the limit case of the conformal galilean algebra $\mbox{\sc cga}(1)$, including the derivation of
the required regularity property.} 
Appendix~B contains mathematical background on Hardy spaces in restricted geometries, for both $d=1$ and $d=2$.

\section{Regularised meta-conformal correlator: the $1D$ case}

Non-holomorphic correlators can only be found by going beyond the local differential operators derived from the
meta-conformal Ward identities. We shall do so in a few simple steps \cite{Henkel16}, 
restricting in this section to the $1D$ case. 
First, we consider the `rapidity' $\gamma$ as a new variable. 
Second, it is dualised \cite{Henkel14a,Henkel15,Henkel15a} through a Fourier transformation, 
which gives the quasi-primary scaling operator
\BEQ \label{Fouriergamma}
\wht{\vph}(\zeta, t, r)=\frac{1}{\sqrt{2\pi}}\int_{\mathbb{R}}\!\D\gamma\; e^{\II \gamma \zeta}\,\vph_{\gamma}(t,r)
\EEQ
This leads to the following representation of the dualised meta-conformal algebra 
\BLAU{(using (\ref{3}) from section~1 with $X_n=\ell_n+\bar{\ell}_n$ and $Y_n=\bar{\ell}_n$ \cite{Henkel02,Henkel19a})}
\BEA
X_n &=& \frac{\II(n+1)}{\mu}\left[\left(t+\mu r\right)^n - t^n\right]\partial_{\zeta} 
-t^{n+1}\partial_t -\frac{1}{\mu}\left[\left(t+\mu r\right)^{n+1}-t^{n+1}\right]\partial_r - (n+1)\delta t^n 
\nonumber \\
Y_n &=& \frac{\II(n+1)}{\mu}\left(t+\mu r\right)^n\partial_{\zeta} -\frac{1}{\mu}\left(t+\mu r\right)^{n+1}\partial_r 
\label{dualvarconf}
\EEA
such that the $1D$ meta-conformal Lie algebra is given by
\BEQ \label{Liemc}
\left[ X_n, X_{m} \right] = (n-m) X_{n+m} \;\; , \;\;
\left[ X_n, Y_{m} \right] = (n-m) Y_{n+m} \;\; , \;\;
\left[ Y_n, Y_{m} \right] = (n-m) Y_{n+m} 
\EEQ
This form will be more convenient for us than the one used in \cite{Henkel16}, since the parameter $\mu$ does no longer appear
in the Lie algebra commutators (\ref{Liemc}). 
Third, it was suggested \cite{Henkel03a,Henkel14a,Henkel16} to look for a further generator $N$ in the Cartan sub-algebra 
$\mathfrak{h}$, viz. $\mbox{\rm ad}_N {\cal X} = \alpha_{\cal X} {\cal X}$ for any meta-conformal generator ${\cal X}$
and $\alpha_{\cal X}\in\mathbb{C}$. 
It can be shown that
\BEQ
N = -\zeta\partial_{\zeta} - r\partial_r + \mu\partial_{\mu} + \II\kappa(\mu) \partial_{\zeta} - \nu(\mu)
\EEQ
is the only possibility \cite{Henkel16}, where the functions $\kappa(\mu)$ and $\nu(\mu)$ remain undetermined. 
Since in this generator, the parameter $\mu$ is treated as a further variable, we see the usefulness of the chosen
normalisation of the generators in (\ref{dualvarconf}). On the other hand, the generator of spatial translations now
reads $Y_{-1}=-\mu^{-1}\partial_r$, with immediate consequences for the form of the two-point correlator. In dual space,
the two-point correlator is defined as
\BEQ
\wht{F} = \left\langle \wht{\vph}_1(\zeta_1,t_1,r_1,\mu_1)\wht{\vph}_2(\zeta_2,t_2,r_2,\mu_2)\right\rangle
= \wht{F}(\zeta_1,\zeta_2,t_1,t_2,r_1,r_2,\mu_1,\mu_2)
\EEQ
Lifting the generators from the representation (\ref{dualvarconf}) to two-body operators, the global 
meta-conformal Ward identities (derived from the maximal finite dimensional sub-algebra isomorphic to
$\mathfrak{sl}(2,\mathbb{C})\oplus\mathfrak{sl}(2,\mathbb{C})$)
become a set of linear partial differential equations of first order ${\cal X}^{[2]}\wht{F}=0$ 
for the function $\wht{F}$. While the solution will
certainly be holomorphic in its (dual) variables, the back-transformation according to (\ref{Fouriergamma}) 
will lead to a correlator bounded everywhere but can introduce non-holomorphic behaviour in the $t_i, r_i$.  

The function $\wht{F}$ is obtained as follows \cite{Henkel16,Henkel19a}. First, co-variance under $X_{-1}$ and $Y_{-1}$ gives
\BEQ
\wht{F} = \wht{F}(\zeta_1,\zeta_2,t,\xi,\mu_1,\mu_2) \;\; ; \;\; t = t_1 - t_2 \;\; , \;\;
\xi = \mu_1 r_1 - \mu_2 r_2
\EEQ
The action of the generators $Y_0$ and $Y_1$ on $\wht{F}$ is used best by introducing the new variables
$\eta := \mu_1\zeta_1 + \mu_2 \zeta_2$ and $\zeta := \mu_1\zeta_1 - \mu_2\zeta_2$. Then the corresponding Ward identities
become
\BEQ
\left( 2\II\partial_{\eta} - (t+\xi)\partial_{\xi} \right) \wht{F} = 0 \;\; , \;\;
\partial_{\zeta} \wht{F}=0
\EEQ
Finally, the Ward identities coming from the generators $X_0$ and $X_1$ become
\BEQ
\left( - t\partial_t - \xi\partial_{\xi} -\delta_1 - \delta_2 \right)\wht{F}=0 \;\; , \;\;
t \left( \delta_1 - \delta_2\right) \wht{F} = 0
\EEQ
The second of these gives the constraint $\delta_1=\delta_2$. The two remaining equations have the general solution
\BEQ \label{FHut-final}
\wht{F} = (t_1-t_2)^{-2\delta_1} \wht{\mathscr{F}}\left( \demi\left(\mu_1\zeta_1+\mu_2\zeta_2\right) 
+ \II\ln\left( 1+ \frac{\mu_1 r_1 - \mu_2 r_2}{t_1 - t_2}\right); \mu_1, \mu_2\right) 
\EEQ
with an undetermined function $\wht{\mathscr{F}}$. 
Spatial translation-invariance only holds in a more weak form, which could become useful for the description of 
physical situations where the propagation speed of each scaling operator can be different.\footnote{See \cite{Moosavi19} for
inhomogeneous ortho-conformal invariance without translation-invariance.} 

In \cite{Henkel16}, we tried to use co-variance under the further generator $N$ in order to fix the function 
$\wht{\mathscr{F}}$, \BLAU{in close analogy with the conformal galilean algebra, see appendix~A.} 
However, therein a choice of basis in the meta-conformal Lie algebra was used where the parameter $\mu$
appears in the structure constants. In this way,  $\wht{\mathscr{F}}$ is fixed and furthermore one can show that
$\wht{F}$ with respect to the variable $\eta$ is in the Hardy space $H_2^+$, see \BLAU{appendix~B} for the mathematical details. 
If we want to consider $\mu$ as a further variable, as it is necessary because of the
explicit form of $N$, objects such as ``$\mu Y_{n+m}$'' 
which arise in the commutators are not part of the meta-conformal Lie algebra. Therefore, it is
necessary, to use the normalisation (\ref{dualvarconf}) which leads to the Lie algebra (\ref{Liemc}) which is independent
of $\mu$. In order to illustrate the generic consequences, let $\nu=\nu(\mu)$ and $\sigma=-\mu\kappa(\mu)$ be constants 
(hence independent of $\mu_{1,2}$). The 
co-variance condition $N\wht{F}=0$ gives 
\BEQ
\wht{\mathscr{F}}(w;\mu_1,\mu_2) 
= \left( \mu_1 \mu_2\right)^{\nu} \wht{\mathpzc{F}}\left(w + \II\sigma\frac{\mu_1+\mu_2}{2}, \frac{\mu_1}{\mu_2}\right)
\EEQ
where the function $\wht{\mathpzc{F}}$ remains undetermined. In contrast to our earlier treatment, we can no longer show
that $\wht{\mathscr{F}}$ had to be in the Hardy space $H_2^+$. On the other hand, this mathematical property had turned out
to be very useful for the derivation of bounded correlators, \BLAU{at least in the conformal galilean case}. This motivates the following. 

First, return to the result (\ref{FHut-final}), rewritten as follows (with the constraint $\delta_1=\delta_2$)
\BEQ \label{17}
\wht{F} = (t_1-t_2)^{-2\delta_1} \wht{\mathscr{F}}(\zeta_+ + \II \lambda) \;\; , \;\;
\zeta_+ := \frac{\mu_1\zeta_1 + \mu_2\zeta_2}{2} \;\; , \;\;
\lambda := \ln\left( 1 + \frac{\mu_1 r_1 - \mu_2 r_2}{t_1-t_2} \right)
\EEQ
and we denote $\wht{\mathscr{F}}_{\lambda}(\zeta_+) := \wht{\mathscr{F}}(\zeta_+ +\II \lambda)$. 
Then, we require (see \BLAU{appendix~B} for details):  \\

\noindent
{\bf Postulate.} {\it If $\lambda>0$, then $\wht{\mathscr{F}}_{\lambda}\in H_2^+$ and if $\lambda<0$, 
then $\wht{\mathscr{F}}_{\lambda}\in H_2^-$.}\\

\noindent
The Hardy spaces $H_2^{\pm}$ on the upper and lower complex half-planes $\mathbb{H}_{\pm}$ are defined in  \BLAU{appendix~B}. 
There, it is also shown that, under mild conditions, that if $\lambda>0$ and if there exist finite positive constants 
$\wht{\mathscr{F}}^{(0)}>0$, $\vep>0$ such that
$|\wht{\mathscr{F}}(\zeta_+ +\II \lambda)| <\wht{\mathscr{F}}^{(0)} e^{-\vep \lambda}$, then $\wht{\mathscr{F}}_\lambda$
is indeed in the Hardy space $H_2^{+}$. Physically, this amounts to a requirement that the dual correlator should
decay algebraically, viz. 
$|\wht{\mathscr{F}}| \leq \wht{\mathscr{F}}_0 \left| 1 + \frac{\mu_1 r_1 - \mu_2 r_2}{t_1-t_2} \right|^{-\vep}$,
with respect to the scaling variable. 
The above postulate also appears natural since in the $\mu\to 0$ limit of conformal galilean invariance, 
it can be derived from the condition of co-variance under the extra generator $N$ \cite{Henkel14a,Henkel15,Henkel15a}. 

The utility of our postulate is easily verified, following \cite{Henkel16}. From Theorem~1 of \BLAU{appendix~B}, especially 
(\ref{A3}), we can write
\BEQ
\wht{\mathscr{F}}_{\lambda}(\zeta_+) = 
\frac{\Theta(\lambda)}{\sqrt{2\pi\,}} \int_0^{\infty} \!\D\gamma_+\: 
e^{\II(\zeta_+ +\II\lambda)\gamma_+} \wht{\mathscr{F}}_+(\gamma_+) +
\frac{\Theta(-\lambda)}{\sqrt{2\pi\,}} \int_0^{\infty} \!\D\gamma_-\: 
e^{-\II(\zeta_+ +\II\lambda)\gamma_-} \wht{\mathscr{F}}_-(\gamma_-)
\EEQ
where the Heaviside functions $\Theta(\pm\lambda)$ select the two cases. For $\lambda>0$, we find
\BEA
F &=& \frac{1}{2\pi} \int_{\mathbb{R}^2} \!\D\zeta_1 \D\zeta_2\: e^{-\II\gamma_1\zeta_1 -\II\gamma_2\zeta_2} \wht{F}
\nonumber \\
&=& \frac{1}{(2\pi)^{3/2}} \int_{\mathbb{R}^2} \!\D\zeta_1 \D\zeta_2\; t^{-2\delta_1} 
\int_0^{\infty} \!\D\gamma_+\: e^{-\II\gamma_1\zeta_1-\II\gamma_2\zeta_2} 
e^{\II (\mu_1\zeta_1+\mu_2\zeta_2+2\II\lambda)\gamma_+ /2} \wht{\mathscr{F}}_+(\gamma_+) 
\nonumber \\
&=& \frac{\sqrt{32\pi\,}}{\mu_1\mu_2} t^{-2\delta_1} \int_0^{\infty} \!\D\gamma_+ \: e^{-\lambda\gamma_+} 
\delta\left(\gamma_+ - \frac{2\gamma_1}{\mu_1}\right)\delta\left(\gamma_+ - \frac{2\gamma_2}{\mu_2}\right) 
\wht{\mathscr{F}}_+(\gamma_+) 
\nonumber \\
&=& \frac{\sqrt{32\pi\,}}{\mu_1\mu_2} t^{-2\delta_1} \delta_{\gamma_1/\mu_1,\gamma_2/\mu_2} 
\int_0^{\infty} \!\D\gamma_+ \: e^{-\lambda \gamma_+} \delta\left(\gamma_+ - \frac{2\gamma_1}{\mu_1}\right) 
\wht{\mathscr{F}}_+(\gamma_+) 
\nonumber \\
&=& \mbox{\rm cste. } \delta_{\gamma_1/\mu_1,\gamma_2/\mu_2} (t_1-t_2)^{-2\delta_1} 
\left( 1 + \frac{\mu_1 r_1 - \mu_2 r_2}{t_1-t_2}\right)^{-2\gamma_1/\mu_1} \Theta\left(\frac{\gamma_1}{\mu_1}\right)
\EEA
where the definitions (\ref{17}) were used. Similarly, for $\lambda<0$ we obtain
\BEA
F &=& \frac{1}{2\pi} \int_{\mathbb{R}^2} \!\D\zeta_1 \D\zeta_2\: e^{-\II\gamma_1\zeta_1 -\II\gamma_2\zeta_2} \wht{F}
\nonumber \\
&=& \frac{1}{(2\pi)^{3/2}} \int_{\mathbb{R}^2} \!\D\zeta_1 \D\zeta_2\; t^{-2\delta_1} 
\int_0^{\infty} \!\D\gamma_-\: e^{-\II\gamma_1\zeta_1-\II\gamma_2\zeta_2} 
e^{-\II (\mu_1\zeta_1+\mu_2\zeta_2+2\II\lambda)\gamma_- /2} \wht{\mathscr{F}}_-(\gamma_-) 
\nonumber \\
&=& \frac{\sqrt{32\pi\,}}{\mu_1\mu_2} t^{-2\delta_1} \int_0^{\infty} \!\D\gamma_- \: e^{\lambda\gamma_-} 
\delta\left(\gamma_- + \frac{2\gamma_1}{\mu_1}\right)\delta\left(\gamma_- + \frac{2\gamma_2}{\mu_2}\right) 
\wht{\mathscr{F}}_-(\gamma_-) 
\nonumber \\
&=& \frac{\sqrt{32\pi\,}}{\mu_1\mu_2} t^{-2\delta_1} \delta_{\gamma_1/\mu_1,\gamma_2/\mu_2} 
\int_0^{\infty} \!\D\gamma_- \: e^{\lambda \gamma_-} 
\delta\left(\gamma_- - \left|\frac{2\gamma_1}{\mu_1}\right|\right) \wht{\mathscr{F}}_-(\gamma_-) 
\nonumber \\
&=& \mbox{\rm cste. } \delta_{\gamma_1/\mu_1,\gamma_2/\mu_2} (t_1-t_2)^{-2\delta_1} 
\left( 1 - \frac{\mu_1 r_1 - \mu_2 r_2}{t_1-t_2}\right)^{-2|\gamma_1/\mu_1|} \Theta\left(-\frac{\gamma_1}{\mu_1}\right)
\EEA
Combining these two forms gives our final $1D$ two-point correlator
\BEQ \label{metareg1d}
F = \delta_{\delta_1,\delta_2}\delta_{\gamma_1/\mu_1,\gamma_2/\mu_2}  
\left( 1 + \left| \frac{\mu_1 r_1 - \mu_2 r_2}{t_1-t_2}\right|\right)^{-2|\gamma_1/\mu_1|}
\EEQ
up to normalisation. As illustrated in figure~\ref{fig00}, this is real-valued and bounded in the entire time-space, 
although not a holomorphic function of the time-space coordinates. 

Finally, it appears that our original motivation for allowing the $\mu_j$ to become free variables, is not very strong. 
We might have fixed the $\mu_j$ from the outset, had not included a factor $1/\mu$ into the generators $Y_n$ (such that
the spatial translations are generated by $Y_{-1}=-\partial_r$ and continue immediately with our Postulate. 
Since a consideration of the meta-conformal three-point function shows that $\mu_1=\mu_2=\mu_3$ \cite[chap. 5]{Henkel10}, 
we can then consider $\mu^{-1}$ as an universal velocity.\footnote{In the conformal galilean limit $\mu\to 0$, recover
the  bounded result $F\sim \exp\left( -2 |\gamma_1 r|/t\right)$ \cite{Henkel16}.}  

\section{Regularised meta-conformal correlator: the $2D$ case}

The derivation of the $2D$ meta-conformal correlator starts essentially along the same lines as in the $1D$ case, but is
based now on the generators (\ref{4}). The dualisation is now carried out with respect to the chiral rapidities
$\gamma=\gamma_{\|}-\II\gamma_{\perp}$ and $\bar{\gamma}=\gamma_{\|}+\II\gamma_{\perp}$ and we also use the light-cone
coordinates $z=r_{\|}+\II r_{\perp}$ and $\bar{z}=r_{\|}-\II r_{\perp}$. The dualisation proceeds as follows 
\BEQ \label{Fouriergamma2D}
\wht{\vph}(\zeta, \bar{\zeta},t, \vec{r})=\frac{1}{2\pi}\int_{\mathbb{R}^2}\!\D\gamma\D\bar{\gamma}\; 
e^{\II (\gamma \zeta+\bar{\gamma}\bar{\zeta})}\,\vph_{\gamma,\bar{\gamma}}(t,\vec{r})
\EEQ
Taking the translation generators $A_{-1}, B_{-1}^{\pm}$ into account, we consider the dual correlator
\BEQ
\wht{F} = \wht{F}(\zeta_1,\zeta_2,\bar{\zeta}_1,\bar{\zeta}_2,t,\xi,\bar{\xi},\mu_1,\mu_2)
\EEQ
where we defined the variables
\BEQ
t = t_1 - t_2 \;\; , \;\;
\xi = \mu_1 z_1 - \mu_2 z_2 \;\; 
\bar{\xi} = \mu_1 \bar{z}_1 - \mu_2 \bar{z}_2
\EEQ
In complete analogy with the $1D$ case, we further define the variables
\BEQ
\eta = \mu_1 \zeta_1 + \mu_2 \zeta_2 \;\; , \;\; \bar{\eta} = \mu_1\bar{\zeta}_1 + \mu_2\bar{\zeta}_2
\EEQ
such that the correlator $\wht{F}=\wht{F}(\eta,\bar{\eta},t,\xi,\bar{\xi},\mu_1,\mu_2)$ obeys the equations
\BEQ \label{25}
\left(2\II\partial_{\eta} - (t+\xi)\partial_{\xi}\right)\wht{F}=0 \;\; , \;\;
\left(2\II\partial_{\bar{\eta}} - (t+\bar{\xi})\partial_{\bar{\xi}}\right)\wht{F}=0 \;\; , \;\;
\left(t\partial_t + \xi\partial_{\xi} + \bar{\xi}\partial_{\bar{\xi}} + 2\delta_1 \right)\wht{F}=0
\EEQ
along with the constraint $\delta_1=\delta_2$. The most general solution of this system is promptly obtained. 
However, in order to obtain real-valued results in terms of the physical coordinates $t,r_{\|},r_{\perp}$, we expand
the logarithms and go back to parallel and perpendicular dual coordinates, via  
$\zeta_j=\demi(\zeta_{\|,j}+\II\zeta_{\perp,j})$ and $\bar{\zeta}_j=\demi(\zeta_{\|,j}-\II\zeta_{\perp,j})$. 
Since the coordinates $\xi,\bar{\xi}$ which occur are complex, we need the complex logarithm
\BEQ
\ln \bigl( a + \II b \bigr) = \demi \ln \bigl( a^2 + b^2 \bigr) + \II \arctan \frac{b}{a}
\EEQ
and find (the first step is the solution of eqs.~(\ref{25}))
\BEA 
\wht{F} &=& t^{-2\delta_1} 
\wht{\mathcal{F}}\left(\frac{\eta}{2} + \II \ln( 1 + \xi/t), \frac{\bar{\eta}}{2} + \II \ln(1+\bar{\xi}/t) \right)
\nonumber \\
&=& t^{-2\delta_1} \wht{\mathscr{F}}\left( \frac{\mu}{2}\left(\zeta_{\|,1}+\zeta_{\|,2}\right) + \II \lambda_{\|}, 
\frac{\mu}{2}\left(\zeta_{\perp,1}+\zeta_{\perp,2}\right) + \II \lambda_{\perp}\right)
\nonumber \\
&=& t^{-2\delta_1} \wht{\mathscr{F}}\left( u + \II \lambda_{\|},  \bar{u} + \II \lambda_{\perp}\right)
\label{26}
\EEA 
with the abbreviations ($\bar{u}$ is obtained from $u$ by replacing $\zeta_{\|,j}\mapsto \zeta_{\perp,j}$) 
\BEQ \label{27}
u := \frac{\mu}{2}(\zeta_{\|,1} + \zeta_{\|,2}) \;\; , \;\;
\lambda_{\|} := \demi \ln\left[ \left( 1 + \frac{\mu r_{\|}}{t}\right)^2 + \left(\frac{\mu r_{\perp}}{t}\right)^2\right]
\;\; , \;\;
\lambda_{\perp} := \arctan\frac{\mu r_{\perp}/t}{1+\mu r_{\|}/t}
\EEQ
and we simplified the notation by letting $\mu_1=\mu_2=\mu$ and assumed translation-invariance in time and space. 
As before, we expect that a Hardy space will permit to derive the boundedness, see  \BLAU{appendix~B} for details. Define 
$\wht{\mathscr{F}}_{\lambda_{\|},\lambda_{\perp}}(u,\bar{u}) 
:=\wht{\mathscr{F}}(u+\II \lambda_{\|},\bar{u}+\II \lambda_{\perp})$ and require: \\

\noindent
\BD
\hspace{-2.5truecm}\mbox{{\bf Postulate:} {\it If the parameters}} \left\{ \begin{array}{l} 
\lambda_{\|}>0\mbox{\it ~and~}\lambda_{\perp}>0\mbox{\it , then $\wht{\mathscr{F}}_{\lambda_{\|},\lambda_{\perp}}\in H_2^{++}$} \\
\lambda_{\|}>0\mbox{\it ~and~}\lambda_{\perp}<0\mbox{\it , then $\wht{\mathscr{F}}_{\lambda_{\|},\lambda_{\perp}}\in H_2^{+-}$} \\
\lambda_{\|}<0\mbox{\it ~and~}\lambda_{\perp}>0\mbox{\it , then $\wht{\mathscr{F}}_{\lambda_{\|},\lambda_{\perp}}\in H_2^{-+}$} \\
\lambda_{\|}<0\mbox{\it ~and~}\lambda_{\perp}<0\mbox{\it , then $\wht{\mathscr{F}}_{\lambda_{\|},\lambda_{\perp}}\in H_2^{--}$} 
\end{array} \right.   ~~~~~~~
\ED
The important point is that these postulates are made with respect to the parallel and perpendicular dual coordinates
$\zeta_{\|}=\zeta+\bar{\zeta}$ and $\zeta_{\perp}=\frac{1}{\II}(\zeta-\bar{\zeta})$. 

Theorem~2 in \BLAU{appendix~B}, especially (\ref{A11}), then states that 
\BEA
\wht{\mathscr{F}}_{\lambda_{\|},\lambda_{\perp}}(u,\bar{u}) &=& \frac{\Theta(\lambda_{\|}) \Theta(\lambda_{\perp})}{2\pi} 
\int_0^{\infty}\!\D \tau\int_0^{\infty}\!\D\bar{\tau}\; 
e^{\II(u+\II\lambda_{\|})\tau +\II(\bar{u}+\II\lambda_{\perp})\bar{\tau}} \wht{\mathscr{F}}_{++}(\tau,\bar{\tau}) \nonumber \\ 
& & + \frac{\Theta(\lambda_{\|}) \Theta(-\lambda_{\perp})}{2\pi} 
\int_0^{\infty}\!\D \tau\int_0^{\infty}\!\D\bar{\tau}\; 
e^{\II(u+\II\lambda_{\|})\tau -\II(\bar{u}+\II\lambda_{\perp})\bar{\tau}} \wht{\mathscr{F}}_{+-}(\tau,\bar{\tau}) 
\nonumber \\ 
& & + \frac{\Theta(-\lambda_{\|}) \Theta(\lambda_{\perp})}{2\pi} 
\int_0^{\infty}\!\D \tau\int_0^{\infty}\!\D\bar{\tau}\; 
e^{-\II(u+\II\lambda_{\|})\tau +\II(\bar{u}+\II\lambda_{\perp})\bar{\tau}} \wht{\mathscr{F}}_{-+}(\tau,\bar{\tau}) 
\nonumber \\ 
& & + \frac{\Theta(-\lambda_{\|}) \Theta(-\lambda_{\perp})}{2\pi} 
\int_0^{\infty}\!\D \tau\int_0^{\infty}\!\D\bar{\tau}\; 
e^{-\II(u+\II\lambda_{\|})\tau -\II(\bar{u}+\II\lambda_{\perp})\bar{\tau}} \wht{\mathscr{F}}_{--}(\tau,\bar{\tau}) 
\EEA
For example, we can write the two-point function in the case $\lambda_{\|}>0$ and $\lambda_{\perp}>0$, with the short-hand 
${\cal D}\zeta := \D\zeta_{\|,1}\D\zeta_{\perp,1}\D\zeta_{\|,2}\D\zeta_{\perp,2}$ and the abbreviations from (\ref{27})
\BEA
F &=& \frac{1}{(2\pi)^2} \int_{\mathbb{R}^4} \!{\cal D}\zeta\;  
e^{-\II\gamma_{\|,1}\zeta_{\|,1}-\II\gamma_{\perp,1}\zeta_{\perp,1}
-\II\gamma_{\|,2}\zeta_{\|,2}-\II\gamma_{\perp,2}\zeta_{\perp,2} }\wht{F}
\nonumber \\
&=& \frac{t^{-2\delta_1}}{(2\pi)^3} \int_{\mathbb{R}^4} \!{\cal D}\zeta\;   
e^{-\II\gamma_{\|,1}\zeta_{\|,1}-\II\gamma_{\perp,1}\zeta_{\perp,1}
-\II\gamma_{\|,2}\zeta_{\|,2}-\II\gamma_{\perp,2}\zeta_{\perp,2}} \times 
\nonumber \\
& & \times
\int_0^{\infty} \!\D\tau \int_0^{\infty} \!\D\bar{\tau}\; 
e^{\II(\mu(\zeta_{\|,1}+\zeta_{\|,2}))\tau/2 - \lambda_{\|} \tau} 
e^{\II(\mu(\zeta_{\perp,1}+\zeta_{\perp,2}))\bar{\tau}/2 -\lambda_{\perp}\bar{\tau}} \wht{\mathscr{F}}_{++}(\tau,\bar{\tau})
\nonumber \\
&=& \frac{t^{-2\delta_1}}{(2\pi)^3} 
\int_0^{\infty} \!\D\tau \int_0^{\infty} \!\D\bar{\tau}\; \wht{\mathscr{F}}_{++}(\tau,\bar{\tau})\: 
e^{-\lambda_{\|}\tau-\lambda_{\perp}\bar{\tau}} \times \nonumber \\
& & \times 
\int_{\mathbb{R}^4} \!{\cal D}\zeta\; 
e^{\II(\mu\tau/2-\gamma_{\|,1})\zeta_{\|,1}}\, e^{\II(\mu\tau/2-\gamma_{\|,2})\zeta_{\|,2}}\, 
e^{\II(\mu\bar{\tau}/2-\gamma_{\perp,1})\zeta_{\perp,1}}\, e^{\II(\mu\bar{\tau}/2-\gamma_{\perp,2})\zeta_{\perp,2}} 
\nonumber \\
&=& \frac{t^{-2\delta_1}}{(2\pi)^3} 
\int_0^{\infty} \!\D\tau \int_0^{\infty} \!\D\bar{\tau}\; \wht{\mathscr{F}}_{++}(\tau,\bar{\tau})\: 
e^{-\lambda_{\|}\tau-\lambda_{\perp}\bar{\tau}} \times \nonumber \\
& & \times 
\int_{\mathbb{R}^4} \!{\cal D}\zeta\; 
e^{\II(\mu\tau-\gamma_{\|,1}-\gamma_{\|,2})\zeta_{\|,+}}\, e^{-\II(\gamma_{\|,1}-\gamma_{\|,2})\zeta_{\|,-}}\, 
e^{\II(\mu\bar{\tau}-\gamma_{\perp,1}-\gamma_{\perp,2})\zeta_{\perp,+}}\, 
e^{-\II(\gamma_{\perp,1}-\gamma_{\perp,2})\zeta_{\perp,-}} 
 \\
&=& \mbox{\rm cste. } t^{-2\delta_1} \delta_{\gamma_{\|,1}/\mu,\gamma_{\|,2}/\mu} 
\delta_{\gamma_{\perp,1}/\mu,\gamma_{\perp,2}/\mu}\: 
\exp\left[-2 \lambda_{\|}\frac{\gamma_{\|,1}}{\mu} - 2\lambda_{\perp}\frac{\gamma_{\perp,1}}{\mu}\right] 
\Theta\left(\frac{\gamma_{\|,1}}{\mu}\right) \Theta\left(\frac{\gamma_{\perp,1}}{\mu}\right)
\nonumber
\EEA
Herein, variables were changed according to $\zeta_{\|,1} = \zeta_{\|,+} + \zeta_{\|,-}$ and 
$\zeta_{\|,2} = \zeta_{\|,+}-\zeta_{\|,-}$ and similarly for the ${\zeta}_{\perp,j}$. 
The other cases are treated in the same manner, in complete analogy with the $1D$ situation. For example, if $\lambda_{\|}<0$ 
and $\lambda_{\perp}<0$, we find  
\BEA
F 
&=& \frac{t^{-2\delta_1}}{(2\pi)^3} \int_{\mathbb{R}^4} \!{\cal D}\zeta\;   
e^{-\II\gamma_{\|,1}\zeta_{\|,1}-\II\gamma_{\perp,1}\zeta_{\perp,1}
-\II\gamma_{\|,2}\zeta_{\|,2}-\II\gamma_{\perp,2}\zeta_{\perp,2}} \times 
\nonumber \\
& & \times
\int_0^{\infty} \!\D\tau \int_0^{\infty} \!\D\bar{\tau}\; 
e^{-\II(\mu(\zeta_{\|,1}+\zeta_{\|,2}))\tau/2 + \lambda_{\|} \tau} 
e^{-\II(\mu(\zeta_{\perp,1}+\zeta_{\perp,2}))\bar{\tau}/2 +\lambda_{\perp}\bar{\tau}} \wht{\mathscr{F}}_{--}(\tau,\bar{\tau})
 \\
&=& \mbox{\rm cste. } t^{-2\delta_1} \delta_{\gamma_{\|,1}/\mu,\gamma_{\|,2}/\mu} 
\delta_{\gamma_{\perp,1}/\mu,\gamma_{\perp,2}/\mu}\: 
\exp\left[-2 \left|\lambda_{\|}\frac{\gamma_{\|,1}}{\mu}\right| 
- 2\left|\lambda_{\perp}\frac{\gamma_{\perp,1}}{\mu}\right|\right] 
\Theta\left(-\frac{\gamma_{\|,1}}{\mu}\right) \Theta\left(-\frac{\gamma_{\perp,1}}{\mu}\right)
\nonumber
\EEA
We also observe the implied positivity conditions on the $\gamma_{\|,1}$ and $\gamma_{\perp,1}$. 

In order to understand the meaning of these expressions, we return to the physical interpretation of the conditions 
$\lambda_{\|,\perp}>0$, or $\lambda_{\|,\perp}<0$. From (\ref{27}), we see first that $\lambda_{\perp}>0$ if and only if
$\mu r_{\perp}/t \left[ 1 + \mu r_{\|}/t\right]^{-1}>0$. On the other hand, concerning the condition $\lambda_{\|}>0$, 
the most restrictive case occurs for $r_{\perp}=0$. 
Then $\lambda_{\|}>0$ is equivalent to $r_{\|}/t>0$. Summarising, we conclude that
\BEQ \label{final}
F = \delta_{\delta_1,\delta_2}\delta_{\gamma_{\,1},\gamma_{\|,2}} \delta_{\gamma_{\perp,1},\gamma_{\perp,2}}\: t^{-2\delta_1}
\left[\left( 1 +\left|\frac{\mu r_{\|}}{t}\right|\right)^2 
+\left(\frac{\mu r_{\perp}}{t}\right)^2\right]^{-|\gamma_{\|,1}/\mu|}
\exp\left[ - \left| \frac{2\gamma_{\perp,1}}{\mu} \arctan\frac{\mu r_{\perp}/t}{1+\mu r_{\|}/t} \right| \right]
\EEQ
up to normalisation, is the final form for the $2D$ meta-conformally co-variant two-point correlator which is bounded in the
entire time-space with points $(t,r_{\|},r_{\perp})\in\mathbb{R}^3$. 

\begin{figure}[tb]
\begin{center}
\includegraphics[width=.475\hsize]{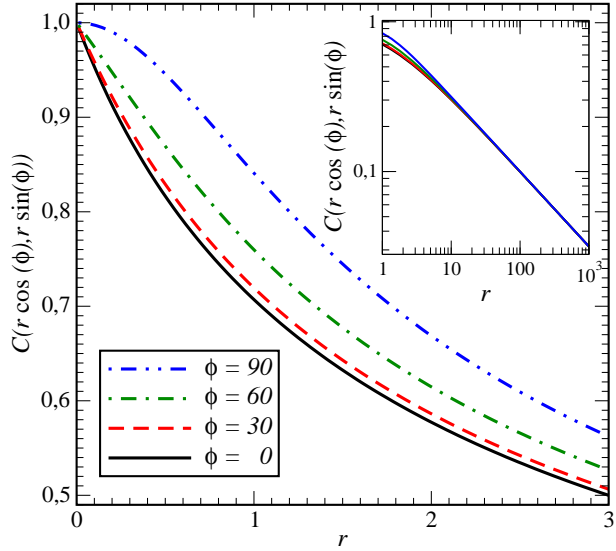} ~~~ 
\end{center}
\caption[fig2]{Shape of the $2D$ meta-conformal two-point correlator 
$C(r\cos\phi,r\sin\phi)= C_{{\rm meta,~} 2D}^{[2]}(1,0,r \cos\phi,0,{r}\sin\phi,0)$ of eq.~(\ref{5.2}), 
with rapidities $\gamma_{\|,1}=\frac{1}{4}$ and $\gamma_{\perp,1}=0$ and with the scales set to
$\mu_1=\mu_2=1$, for the angles $\phi=[0^o, 30^o, 60^o, 90^o]$. The inset shows the algebraic decay 
$C(r)\sim r^{-1/2}$ for large distances. \label{fig2}}
\end{figure}

In figure~\ref{fig2}, we show that the shape of this correlator (\ref{final}) smoothly interpolates 
between the preferred and the transverse direction, when the transverse rapidity $\gamma_{\perp,1}=0$. 
This is qualitatively very similar to the $1D$ meta-conformal correlator, which has been described before 
\cite{Henkel19a,Henkel19b}. For $\phi\ne 90^o$, the cusp at the origin indicates a non-analyticity. 

\begin{figure}[tb]
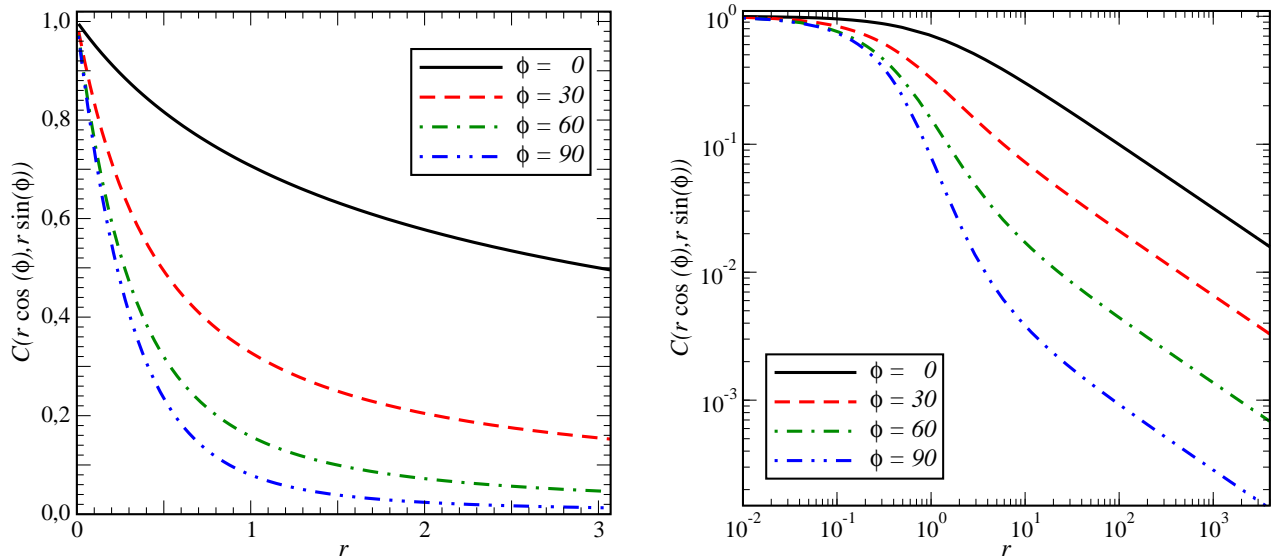

\begin{center}
\includegraphics[width=.475\hsize]{stoimenov13_meta_gamp150_phi.eps} ~~~
\includegraphics[width=.475\hsize]{stoimenov13_meta_gamp150_phi_log.eps} 
\end{center}
\caption[fig3]{Shape of the $2D$ meta-conformal two-point correlator 
$C(r\cos\phi,r\sin\phi)= C_{{\rm meta,~} 2D}^{[2]}(1,0,r \cos\phi,0,{r}\sin\phi,0)$ of eq.~(\ref{5.2}), 
with rapidities $\gamma_{\|,1}=\frac{1}{4}$ and $\gamma_{\perp,1}=\frac{3}{2}$ and with the scales set to
$\mu_1=\mu_2=1$, for the angles $\phi=[0^o, 30^o, 60^o, 90^o]$. The left panel shows the correlator on a linear scale, 
the right panel is on a double logarithmic scale. \label{fig3}}
\end{figure}

A very different behaviour is observed whenever $\gamma_{\perp,1}\ne 0$, as is illustrated in figure~\ref{fig3}.
For all values of $\phi$, there is a cusp at the origin, but the decay for larger values of $r$ is considerably changed with
respect to figure~\ref{fig2}, whenever $\phi\ne 0^o$. Indeed, for a non-vanishing transverse rapidity $\gamma_{\perp,1}\ne 0$,  
the decay with $r$ becomes more fast when the angle $\phi$ increases, 
{\it a contrario} to what is seen in figure~\ref{fig2} where we have $\gamma_{\perp,1}=0$. 
Instead of an algebraic decay, we rather find from figure~\ref{fig3} an exponentially
decaying correlator and only when for very large distances the ratio $r_{\perp}/(1+r_{\|})\to \tan\phi$, there is a cross-over 
to an algebraic decay,\footnote{In the extreme case $\gamma_{\|,1}=0$, the (connected !) correlator $C(r)$ tends towards
a constant when $r\to\infty$.}  $C(r)\sim r^{-2\gamma_{\|,1}/\mu_1}$.

Several further comments are in order: 
\begin{enumerate}
\item Only if the Hardy property is postulated on the parallel and perpendicular dual coordinates $\zeta_{\|}$ and
$\zeta_{\perp}$, does one obtain a completely
bounded expression. Analogous postulates on different coordinates will lead to more weak results (typically for only
one of the parameters $\lambda_{\|}$ or $\lambda_{\perp}$). 
\item the $1D$ case is contained in (\ref{final}) as the special case $r_{\perp}=0$. 
\item in the limit $\mu\to 0$, one recovers the expected result of the $2D$ conformal galilean algebra
\BEA
C_{\mbox{\sc\footnotesize cga}}^{[2]}(t,r_{\|},r_{\perp}) &=& 
\lim_{\mu\to 0} C_{\rm meta}^{[2]}(t,r_{\|},r_{\perp}) \nonumber \\
&=& 
\delta_{\delta_1,\delta_2}\delta_{\gamma_{\,1},\gamma_{\|,2}} \delta_{\gamma_{\perp,1},\gamma_{\perp,2}}\: t^{-2\delta_1}
\exp\left( - 2 \left|\frac{\gamma_{\|,1} r_{\|}}{t}\right| - 2 \left|\frac{\gamma_{\perp,1} r_{\perp}}{t}\right| \right)
\EEA
which could be recast into a vectorial form, as in (\ref{1.5}), but now the correlator is bounded everywhere.  
\item the special case $r_{\|}=0$ reproduces the (non-unitary) {\em ortho}-conformal correlator (\ref{1.6}), 
with the identifications $\gamma=\gamma_{\perp,1}$ and $\delta=|\gamma_{\|,1}/\mu|=\delta_1$.
\end{enumerate}

\section{Conclusions}

Having raised the question how to formulate sufficient criteria such that the meta-conformally co-variant two-point
functions remain bounded in the entire time-space, we have shown that a refined form
of the global Ward identities is needed. These are formulated in a dual space, where the dual variables are naturally
confined to a tube on a half-space for $d=1$ or on one of the quadrants of the dual plane for $d=2$. Then the regularity
condition, namely that these dual two-point functions belong to a certain Hardy space, is a sufficient condition for
the construction of bounded two-point functions. This leads to 
\BEQ \label{5.1}
C_{{\rm meta,~} 1D}^{[2]}(t_1,t_2,r_1,r_2) 
= \delta_{\delta_1,\delta_2}\delta_{\gamma_1/\mu_1,\gamma_2/\mu_2}\:  (t_1-t_2)^{-2\delta_1} 
\left( 1 + \left| \frac{\mu_1 r_1 - \mu_2 r_2}{t_1-t_2}\right|\right)^{-2|\gamma_1/\mu_1|}
\EEQ
(up to normalisation) in $d=1$ spatial dimensions, see also figure~\ref{fig00}, 
and 
\BEA
\lefteqn{\hspace{-1.0truecm}C_{{\rm meta,~} 2D}^{[2]}(t_1,t_2,r_{\|,1},r_{\|,2},{r}_{\perp,1},{r}_{\perp,2}) 
= \delta_{\delta_1,\delta_2}\delta_{\gamma_{1,\|/\mu_1},\gamma_{2,\|}/\mu_2}
\delta_{\gamma_{\perp,1}/\mu_1,\gamma_{\perp,2}/\mu_2} \: (t_1-t_2)^{-2\delta_1} \times~~~~~~} ~~
\nonumber \\
&\times & \left[\left( 1 +\left|\frac{\mu_1 r_{\|,1}-\mu_2 r_{\|,2} }{t_1-t_2}\right|\right)^2 
+\left(\frac{\mu_1 {r}_{\perp,1}-\mu_2 {r}_{\perp,2} }{t_1-t_2}\right)^2\right]^{-|\gamma_{1,\|}/\mu_1|} \times
\nonumber \\
&\times & 
\exp\left[ - \left| \frac{2\gamma_{\perp,1}}{\mu_1} 
\arctan\frac{(\mu_1 r_{\perp,1}-\mu_2r_{\perp,2})/(t_1-t_2)}{1+(\mu_1 r_{\|,1}-\mu_2 r_{\|,2})/(t_1-t_2)} \right| \right]
\label{5.2}
\EEA
in $d=2$ spatial dimensions. We also see that the parallel and perpendicular 
rapidities $\gamma_{\|,1}$ and $\gamma_{\perp,1}$, respectively, 
enter into the $2D$ two-point correlator in very different ways. 
The qualitative shapes of the correlator have been shown in figures~\ref{fig2} and~\ref{fig3} which illustrate clearly
the respective importance of the longitudinal and transverse rapidities, $\gamma_{\|}$ and $\gamma_{\perp}$, together 
with the richness of possible scaling forms, in the real physical coordinates. 
In general, one would have expected that rotation-invariance in the transverse space coordinates 
$\vec{r}_{\perp}$ should reduce the case of $d\geq 3$ dimensions to (\ref{5.2}). 
Surprisingly, this only seems to work\footnote{There is no obvious
rotation-invariant generalisation of the last exponential factor in (\ref{5.2}), beyond a single transverse dimension.} 
if the transverse rapidity $\gamma_{\perp,1}=0$,
when the two-point correlator takes indeed a form which is rotation-invariant in the $d-1$ transverse directions. 
In that case, one may replace $r_{\perp,j} \mapsto \vec{r}_{\perp,j}$.

These explicit results clearly show that boundedness on the entire set of time-space points 
$(t,r_{\|},r_{\perp})\in\mathbb{R}^3$ is not compatible with a holomorphic dependence of the two-point function
on the time- and space-coordinates. According to the Wiener-Khintchine theorem \cite{Fell71}, 
the Fourier transform of a two-point
correlator which obeys spatial translation-invariance must be positive. 
We have checked that eqs.~(\ref{5.1},\ref{5.2}) obey this necessary condition for a physical correlator. 

The known regularised two-point correlators of the conformal galilean algebra \cite{Henkel15,Henkel16}, \BLAU{see eq.~(\ref{1.8}),}  
are reproduced in the $\mu\to 0$ limit. 

Remarkably, questions of causality and boundedness appear to find different answers for the non-semi-simple
Schr\"odinger and conformal galilean algebras than for the semi-simple conformal and meta-conformal algebras, 
see table~\ref{tab1}. 
In the first case, it turned out to be sufficient to enlarge the Cartan sub-algebra $\mathfrak{h}$ 
of the Lie algebra by a further generator $N$. This either immediately yielded the physically required 
causality condition of the response functions whose form is described by the Schr\"odinger algebras 
\cite{Henkel03a,Henkel14a}, or made it possible to demonstrate that the  dualised correlator is in the Hardy space 
$H_2^{\pm}$ \BLAU{(if $d=1$)} \cite{Henkel15,Henkel15a,Henkel16}, \BLAU{see appendix~A}. Although the extra generator $N$ ceases to be useful for the meta-conformal
algebra, the Hardy-space criterion does remain useful, even as a postulate. Of course, for 
{\em ortho}-conformal representations the implicit analyticity in the time-space coordinates permits to sidestep the question. 
\BLAU{It would be interesting to extend these results to the logarithmic representations \cite{Henkel14b} of conformal galilean and meta-conformal algebras.}

\newpage

\appsection{A}{Bounded correlators for the conformal galilean algebra}
\BLAU{To make this work more self-contained, we briefly recall the corresponding argument for the limit case
$\mu\to 0$ of the conformal galilean algebra $\mbox{\sc cga}(1)$ \cite{Henkel15,Henkel15a,Henkel16}. In this case, the Hardy-space property of 
the dualised correlator can be derived and need not be postulated. In order to carry out the required contraction, we start 
from the $1D$ meta-conformal generators (\ref{dualvarconf}) and rescale the generators $\mu Y_n \mapsto Y_n$. 
This changes the commutators (\ref{Liemc}) into 
\BEQ 
\left[ X_n, X_{m} \right] = (n-m) X_{n+m} \;\; , \;\;
\left[ X_n, Y_{m} \right] = (n-m) Y_{n+m} \;\; , \;\;
\left[ Y_n, Y_{m} \right] = (n-m) \mu Y_{n+m} 
\EEQ
Taking now the limit $\mu\to 0$ in (\ref{dualvarconf}) gives the contracted generators
\BEA
X_n &=& \II(n+1)n t^{n-1} r\partial_{\zeta} -t^{n+1}\partial_t -(n+1) t^{n} r\partial_r - (n+1)\delta t^n 
\nonumber \\
Y_n &=& {\II(n+1)} t^n\partial_{\zeta} -t^{n+1}\partial_r 
\EEA
of the non-semi-simple conformal galilean algebra $\mbox{\sc cga}(1)$, with the commutators 
\BEQ 
\left[ X_n, X_{m} \right] = (n-m) X_{n+m} \;\; , \;\;
\left[ X_n, Y_{m} \right] = (n-m) Y_{n+m} \;\; , \;\;
\left[ Y_n, Y_{m} \right] = 0
\EEQ
}

\BLAU{We proceed as in section~2 of the main text. Add an extra generator $N$ to the Cartan subalgebra, which must read 
\BEQ
N = -\zeta\partial_{\zeta} - r\partial_r - \nu
\EEQ
where $\nu$ is a constant. As in section~2 of the main text, the dualised two-point correlator must take the form 
$\wht{F} = \delta_{\delta_1,\delta_2}\,|t|^{-2\delta_1}  \wht{\mathscr{F}}\left(\zeta_+ +\II r/t\right)$. 
Recall the notation $\wht{\mathscr{F}}_{\lambda}(\zeta_+) := \wht{\mathscr{F}}(\zeta_+ +\II \lambda)$ and $\zeta_+=\demi\left(\zeta_1+\zeta_2\right)$. 
In the case at hand co-variance under $N$ leads to the specific form \cite{Henkel15,Henkel15a}
\BEQ
\wht{\mathscr{F}}(u) = u^{-2\nu} \;\; , \;\; 2\nu := \nu_1 + \nu_2
\EEQ
up to an unspecified normalisation constant. We can now make contact with the Hardy spaces $H_2^{\pm}$ introduced in appendix~B.}

\noindent
\BLAU{{\bf Proposition} \cite{Henkel15,Henkel15a}. {\it Let $\nu>\frac{1}{4}$. If $\lambda>0$, then $\wht{\mathscr{F}}_{\lambda}\in H_2^+$ 
and if $\lambda<0$, then $\wht{\mathscr{F}}_{\lambda}\in H_2^-$.}} 

\noindent
\BLAU{{\bf Proof}: $\wht{\mathscr{F}}_{\lambda}$ is clearly analytic in the complex half-planes $\mathbb{H}_{\pm}$, 
respectively and we must check the bound (\ref{B:Hardy1d}). Observe that 
$\left|\wht{\mathscr{F}}_{\lambda}(u+\II v)\right| = \left| \left( u +\II(v+\lambda) \right)^{-2\nu} \right| = \left( u^2 + (v+\lambda)^2 \right)^{-\nu}$. 
Since $\nu>\frac{1}{4}$ the integral in (\ref{B:Hardy1d}) converges and we have explicitly, for $\lambda>0$
\BD
M^2 = \sup_{{v}>0} \int_{\mathbb{R}} \!\D u\: \left| \wht{\mathscr{F}}_{\lambda}(u+\II {v})\right|^2 
=  \frac{\sqrt{\pi\,}\: \Gamma(2\nu-\demi)}{\Gamma(2\nu)} 
\sup_{{v}>0} \left({v}+\lambda\right)^{1-4\nu}
< \infty
\ED
as required ($\Gamma$ denotes Euler's Gamma function \cite{Abra65}). For $\lambda<0$ the argument is analogous. \hfill \qed}

\BLAU{The final correlator is found from the Hardy space representation theorem (\ref{A3}). For $\lambda>0$, this gives (with the constraint $\delta_1=\delta_2$)
\BEA
F &=& \frac{|t|^{-2x_1}}{\pi\sqrt{2\pi}} \int_{\mathbb{R}^2} \!\D\zeta_+ \D\zeta_-\: 
e^{-\II(\gamma_1+\gamma_2)\zeta_+} e^{-\II(\gamma_1-\gamma_2)\zeta_{-}} 
\int_{\mathbb{R}} \!\D\gamma_+\: 
\Theta(\gamma_+)\wht{{\cal F}_+}(\gamma_+) e^{-\gamma_+ \lambda} e^{\II\gamma_+\zeta_+}
\nonumber \\
&=& \frac{|t|^{-2x_1}}{\pi\sqrt{2\pi}} 
\int_{\mathbb{R}} \!\D\gamma_+\:  \Theta(\gamma_+)\wht{{\cal F}_+}(\gamma_+) e^{-\gamma_+ \lambda}
\int_{\mathbb{R}}\!\D\zeta_{-}\: e^{-\II(\gamma_1-\gamma_2)\zeta_{-}}
\int_{\mathbb{R}}\!\D\zeta_{+}\: e^{\II(\gamma_+-\gamma_1-\gamma_2)\zeta_+}
\nonumber \\
&=& \delta(\gamma_1-\gamma_2) \Theta(\gamma_1) F_{0,+}(\gamma_1) e^{-2\gamma_1 \lambda} |t|^{-2\delta_1}
\EEA 
Herein, $F_{0,+}$ contains the unspecified dependence on 
the positive constant $\gamma_1$. Similar, for $\lambda<0$, we find 
$F = \delta(\gamma_1-\gamma_2) \Theta(-\gamma_1) F_{0,-}(\gamma_1) e^{2\gamma_1 |\lambda|} |t|^{-2\delta_1}$.} 

\BLAU{Before writing down a single combined form, we generalise to $\mbox{\sc gca}(d)$ in $d>1$ dimensions. 
Using a vector notation, the dualised generators are (with $n\in\mathbb{Z}$, and $1\leq j,k\leq d$) \cite{Henkel15} 
\BEA
X_n &=& +\II(n+1)n t^{n-1} \vec{r}\cdot\partial_{\vec{\zeta}} -t^{n+1}\partial_t -(n+1) t^n \vec{r}\cdot \partial_{\vec{r}}  -(n+1) \delta t^n
\nonumber \\
\vec{Y}_n &=& +\II (n+1) t^n \partial_{\vec{\zeta}} -t^{n+1}\partial_{\vec{r}}   \\
R_n^{(jk)} &=& - t^n \left( r_j \partial_{r_k} -  r_k \partial_{r_j} \right) 
- t^n \left( \zeta_j \partial_{\zeta_k} -  \zeta_k \partial_{\zeta_j} \right) \nonumber \\
N &=& -\vec{\zeta}\cdot\partial_{\vec{\zeta}} - \vec{r}\cdot \partial_{\vec{r}} - \nu \nonumber
\EEA
Taking also rotation-invariance (notice the specific form of the dualised spatial rotation generator $R_{0}^{(jk)}$) 
into account, we recover eq.~(\ref{1.8}) of the main text. 
}

\appsection{B}{Background on Hardy spaces}

In the main text \BLAU{(and appendix~A)}, we need precise statements on the Fourier transform on semi-infinite spaces. These can be conveniently
formulated in terms of Hardy spaces, where we restrict to the special case $H_2$. Our brief summary is based on 
\cite{Akhiezer88,Stein71}. 

We begin with the case of functions of a single complex variable $z$, defined in the upper half-plane 
$\mathbb{H}_+ := \left\{ z\in\mathbb{C}\left| z=x+\II y \mbox{\rm ~and } y> 0\right.\right\}$. 

\noindent
{\bf Definition 1:} {\it A function $f:\mathbb{H}_+\to \mathbb{C}$ belongs to the {\em Hardy space $H_2^+$} 
if it is holomorphic on $\mathbb{H}_+$ and if}
\BEQ \label{B:Hardy1d}
M^2 := \sup_{y>0} \int_{-\infty}^{\infty} \!\D x\: \left| f(x+\II y)\right|^2 < \infty
\EEQ
The main results of interest to us can be summarised as follows. 

\noindent
{\bf Theorem 1:} \cite{Akhiezer88} 
{\it Let $f:\mathbb{H}_+\to \mathbb{C}$ be a holomorphic function. Then the following statements are equivalent:
\begin{enumerate}
\item $f\in H_2^+$
\item there exists a function $\mathpzc{f}: \mathbb{R}\to\mathbb{C}$, 
which is square-integrable $\mathpzc{f}\in L^2(\mathbb{R})$,
such that $\lim_{y\to 0^+} f(x+\II y) =  \mathpzc{f}(x)$ and
\BEQ \label{A2}
f(z) = \frac{1}{2\pi \II}\int_{-\infty}^{\infty} \!\D\xi\: \frac{\mathpzc{f}(\xi)}{\xi-z} \;\; , \;\;
0 = \frac{1}{2\pi \II}\int_{-\infty}^{\infty} \!\D\xi\: \frac{\mathpzc{f}(\xi)}{\xi-z^*} 
\EEQ
where $z^*=x-\II y$ denotes the complex conjugate of $z$. For notational simplicity, one often writes
$f(x) = \lim_{y\to 0^+} f(x+\II y)$, with $x\in\mathbb{R}$. 
\item there exists a function $\wht{f}:\mathbb{R}_+\to\mathbb{C}$, $\wht{f}\in L^2( \mathbb{R}_+)$, such that for all $y>0$
\BEQ \label{A3}
f(z) = f(x+\II y) = \frac{1}{\sqrt{2\pi\,}\,} \int_0^{\infty} \!\D \zeta\: e^{\II (x+\II y) \zeta} \wht{f}(\zeta)
\EEQ
\end{enumerate}}
The property (\ref{A3}) is of major interest to us in the main text. 

Clearly, any function $f:\mathbb{H}_+ \to \mathbb{C}$ which also admits a representation (\ref{A3}) with $y>0$ and 
$\wht{f}\in L^2( \mathbb{R}_+)$ is in the Hardy space $H_2^+$, since
\BEA
\int_{\mathbb{R}} \!\D x\: |f(x+\II y)|^2 &=&
\frac{1}{2\pi}\int_0^{\infty}\!\D\zeta \int_0^{\infty}\!\D\bar{\zeta}\: \wht{f}(\zeta)\wht{f}^*(\bar{\zeta})
\int_{\mathbb{R}} \!\D x\: e^{\II(x+\II y)\zeta}\, e^{-\II(x-\II y)\bar{\zeta}} \nonumber \\
&=& \frac{1}{2\pi}\int_0^{\infty}\!\D\zeta \int_0^{\infty}\!\D\bar{\zeta}\: \wht{f}(\zeta)\wht{f}^*(\bar{\zeta}) \,
e^{-2y (\zeta+\bar{\zeta})} \int_{\mathbb{R}} \!\D x\: e^{\II x(\zeta-\bar{\zeta})} \nonumber \\
&=& \int_0^{\infty} \!\D\zeta\: |\wht{f}(\zeta)|^2\, e^{-4y\zeta} 
\:\leq\: \int_0^{\infty} \!\D\zeta\: |\wht{f}(\zeta)|^2 \: =:\: M^2 <\infty
\EEA
The more difficult part is to show that any $f\in H_2^+$ indeed admits such a representation. 
For the proof, (\ref{A2}) is needed \cite{Akhiezer88}. 
Elements of a Hardy space enjoy certain limit behaviours, e.g. if $f\in H_2^+$, it can easily be shown that \cite{Akhiezer88}
\begin{subequations}
\begin{align}
\lim_{y\to \infty} f(x+\II y)=0       & \mbox{\rm ~~;~ uniformly for all $x\in\mathbb{R}$}  \label{ann:limite1}\\
\lim_{x\to \pm \infty} f(x+\II y) = 0 & \mbox{\rm ~~;~ uniformly with respect to $y\geq y_0 >0$}
\end{align}
Indeed, eq.~(\ref{ann:limite1}) follows from the bound (which in turn can be obtained from (\ref{A3}) 
\cite{Akhiezer88,Koosis98})
\begin{align}
|f(x+\II y)| &\leq f_{\infty} y^{-1/2}
\end{align}
\end{subequations}
which holds for all $x\in\mathbb{R}$ and where the constant $f_{\infty}>0$ depends on the function $f$. A simple 
sufficient criterion establishes whether a given function $f$ is in the Hardy space $H_2^+$: 

\noindent 
{\bf Lemma:} {\it If the complex function $f(z)=f(x+\II y)$ is holomorphic for all $y\geq 0$, obeys the bound
$|f(z)| < f_0 e^{-\delta y}$, with constants $f_0>0$ and $\delta>0$, and if 
$\int_{-\infty}^{\infty} \!\D x\: |f(x)|^2< \infty$, then $f\in H_2^+$.} 

\noindent
{\bf Proof:} Since $f(z)$ is holomorphic on the closure $\overline{\mathbb{H}_+}$ (which includes the real axis), 
one has the Cauchy formula
\BD
f(z) = \frac{1}{2\pi \II}\oint_{\mathscr{C}} \!\D w\: \frac{f(w)}{w-z} 
= \frac{1}{2\pi \II}\int_{-R}^{R} \!\D w\: \frac{f(w)}{w-z}  
+ \frac{1}{2\pi \II}\int_{\mathscr{C}_{\rm sup}} \!\D w\: \frac{f(w)}{w-z} =: F_1(z) + F_2(z) 
\ED
where the integration contour $\mathscr{C}$ consists of the segment $[-R,R]$ on the real axis and the superior semi-circle 
$\mathscr{C}_{\rm sup}$. One may write $w=u+\II v = R e^{\II \theta}\in \mathscr{C}_{\rm sup}$. It follows that on the
superior semi-circle $|f(w)| < f_0 e^{-\delta v} = f_0 e^{-\delta R \sin \theta}$. Now, for $R$ large enough, one has
$|w-z|= | w(1-z/w)| \geq R \demi$, for $z\in\overline{\mathbb{H}_+}$ fixed and $w\in\mathscr{C}_{\rm sup}$. One 
estimates the contribution $F_2(z)$ of the superior semi-circle, as follows
\BEA
|F_2(z)| &\leq& \frac{1}{2\pi} \int_{\mathscr{C}_{\rm sup}} \!|\D w| \frac{|f(w)|}{|w (1-z/w)|} 
\:\leq\: \frac{1}{2\pi} \int_0^{\pi} \!\D\theta\: \frac{f_0 e^{-\delta R \sin\theta} R}{R \; \demi} 
\nonumber \\
&\leq & \frac{2 f_0}{\pi} \int_0^{\pi/2} \!\D\theta\: \exp\left( - \frac{2\delta}{\pi} R \theta\right) 
\:\leq\: \frac{f_0}{\delta} \frac{1}{R} \to 0 \mbox{\rm ~~,~ for $R\to \infty$}
\nonumber
\EEA
Hence, for $R\to\infty$, one has the integral representation $f(z) = \frac{1}{2\pi\II}\int_{\mathbb{R}} \!\D w\: 
f(w) (w-z)^{-1}$. Since $f\in L^2(\mathbb{R})$, the assertion follows from eq.~(\ref{A2}) of theorem~1. \hfill \qed 

One may also define a Hardy space $H_2^-$ for functions holomorphic on the lower complex half-plane $\mathbb{H}_-$, 
by adapting the above definition. All results transpose in an evident way. 

Further conceptual preparations are necessary for the generalisation of these results to higher dimensions, 
although we shall merely treat the $2D$ case, which is enough for our purposes (and generalisations to $n>2$ will be obvious). 
We denote $\vec{z}=(z_1,z_2)\in\mathbb{C}^2$ and write the scalar product $\vec{z}\cdot\vec{w}=z_1 w_1 + z_2 w_2$ 
for $\vec{z},\vec{w}\in\mathbb{C}^2$. Following \cite{Stein71}, $H_2$-spaces can be defined as follows.

\noindent
{\bf Definition 2:} {\it If $B\subset \mathbb{R}^2$ is an open set, the {\em tube $T_B$ with base $B$} is}
\BEQ
T_B := \left\{ \vec{z}=\vec{x}+\II \vec{y} \in\mathbb{C}^2 \left| \vec{y}\in B, \vec{x}\in\mathbb{R}^2 \right. \right\}
\EEQ
{\it A function $f: T_B\to\mathbb{C}$ which is holomorphic on $T_B$ is in the {\em Hardy space $H_2(T_B)$} if}
\BEQ
M^2 := \sup_{\vec{y}\in B} \int_{\mathbb{R}^2} \D\vec{x}\: |f(\vec{x}+\II\vec{y})|^2 < \infty
\EEQ
However, it turns out that this definition is too general. More interesting results are obtained if one uses
c\^ones as a base of the tubes. 

\noindent 
{\bf Definition 3:} {\it (i) An {\em open c\^one} $\Gamma\subset\mathbb{R}^n$ satisfies the properties $0\not\in\Gamma$ and
if $\vec{x},\vec{y}\in\Gamma$ and $\alpha,\beta>0$, then $\alpha \vec{x}+\beta\vec{y}\in\Gamma$. A {\em closed c\^one} is the
closure $\overline{\Gamma}$ of an open c\^one $\Gamma$. \\
(ii) If $\Gamma$ is a c\^one, and if the set}
\BEQ
\Gamma^* := \left\{ \vec{x}\in\mathbb{R}^n \left| \vec{x}\cdot \vec{t} \geq 0 
                    \mbox{\rm ~with~} \vec{t}\in\Gamma\right.\right\}
\EEQ
{\it has a non-vanishing interior, then $\Gamma^*$ is the {\em dual c\^one} with respect to $\Gamma$. 
The c\^one $\Gamma$ is called {\em self-dual}, if $\Gamma^* = \overline{\Gamma}$.}

For illustration, note that in one dimension ($n=1$) the only c\^one is 
$\Gamma=\left\{ x\in\mathbb{R}\left| x>0\right.\right\} = \mathbb{R}_+$. 
It is self-dual, since $\Gamma^*=\overline{\Gamma}=\mathbb{R}_{0,+}$. 
In two dimensions ($n=2$), consider the c\^one 
$\Gamma^{++}:= \left\{ \vec{x}\in\mathbb{R}^2\left| \vec{x}=(x_1,x_2) \mbox{\rm ~with~} x_1>0, x_2>0\right.\right\}$ which
is the {\em first quadrant} in the $2D$ plane. Since
\BEQ
{\Gamma^{++\,}}^* 
= \left\{ \vec{x}\in\mathbb{R}^2\left| \vec{x}\cdot\vec{t}\geq 0, \mbox{\rm ~for all~} \vec{t}\in\Gamma^{++}\right.\right\} 
= \mathbb{R}_{0,+} \oplus \mathbb{R}_{0,+} = \overline{\Gamma^{++}}
\EEQ
the c\^one $\Gamma^{++}$ is self-dual. 

Hardy spaces defined on the tubes $T_{\Gamma^{++}}$ of the first quadrant provide the required  structure. 

\noindent
{\bf Definition 4:} \cite{Stein71} {\it If $\Gamma^{++}$ denotes the first quadrant of the plane $\mathbb{R}^2$, 
a function $f: T_{\Gamma^{++}}\to \mathbb{C}$ holomorphic on $T_{\Gamma^{++}}$ is in the {\em Hardy space $H_2^{++}$} if}
\BEQ
M^2 := \sup_{\vec{y}\in \Gamma^{++}} \int_{\mathbb{R}^2} \D\vec{x}\: |f(\vec{x}+\II\vec{y})|^2 < \infty
\EEQ

\noindent
{\bf Theorem 2:} \cite{Stein71} 
{\it Let the function $f:T_{\Gamma^{++}}\to \mathbb{C}$ be holomorphic. Then the following statements are equivalent:
\begin{enumerate}
\item $f\in H_2^{++}$ 
\item there exists a function $\mathpzc{f}: \mathbb{R}^2\to\mathbb{C}$, 
which is square-integrable $\mathpzc{f}\in L^2(\mathbb{R}^2)$,
such that $\lim_{\vec{y}\to \vec{0}^+} f(\vec{x}+\II \vec{y}) =  \mathpzc{f}(\vec{x})$ and
\BEQ 
f(\vec{z}) = \frac{1}{(2\pi \II)^2}\int_{\mathbb{R}^2} \!\D\vec{w}\: \frac{\mathpzc{f}(\vec{w})}{\vec{w}-\vec{z}} \;\; , \;\;
0 = \frac{1}{(2\pi \II)^2}\int_{\mathbb{R}^2} \!\D\vec{w}\: \frac{\mathpzc{f}(\vec{w})}{\vec{w}-\vec{z}^*} 
\EEQ
where $(\vec{w}-\vec{z})^{-1} := (w_1-z_1)^{-1} (w_2-z_2)^{-1}$ and $\vec{z}^*=\vec{x}-\II \vec{y}$ denotes the complex 
conjugate of $\vec{z}$. For notational simplicity, one often writes
$f(\vec{x}) = \lim_{\vec{y}\to \vec{0}^+} f(\vec{x}+\II \vec{y})$, with $\vec{x}\in\mathbb{R}^2$. 
\item there exists a function $\wht{f}:\mathbb{R}_+\oplus\mathbb{R}_+\to\mathbb{C}$, 
with $\wht{f}\in L^2( \mathbb{R}_+\oplus\mathbb{R}_+)$  such that for all $z_i\in\mathbb{H}_+$
\BEQ \label{A11}
f(\vec{z}) = \frac{1}{2\pi} \int_{\Gamma^{++}} \!\D\vec{t}\: e^{\II \vec{z}\cdot\vec{t}} \wht{f}(\vec{t})
= \frac{1}{2\pi} \int_0^{\infty} \!\D t_1 \int_0^{\infty} \!\D t_2\: e^{\II(z_1 t_1 + z_2 t_2)} \wht{f}(\vec{t})
\EEQ
\end{enumerate}
}
The property (\ref{A11}) is of major interest to us in the main text. Summarising, the restriction to the first quadrant
$\Gamma^{++}$ allows to carry over the known results from the $1D$ case, separately for each component. 

Hardy spaces $H_2^{+-}$, $H_2^{-+}$, $H_2^{--}$ on the other quadrants can be defined in complete analogy.

\noindent
{\bf  Acknowledgements:} MH thanks the PHC Rila for support and the MPIPKS Dresden (Germany) for
warm hospitality, where part of this work was done. SS is supported by Bulgarian National
Science Fund Grant KP-06-N28/6.


\end{document}